\documentclass[sigconf]{acmart}
\newcommand{\sys}{CodeTailor}

\AtBeginDocument{%
  \providecommand\BibTeX{{%
    \normalfont B\kern-0.5em{\scshape i\kern-0.25em b}\kern-0.8em\TeX}}}

\copyrightyear{2024}
\acmYear{2024}
\setcopyright{rightsretained}
\acmConference[SIGCSE 2024]{Proceedings of the 55th ACM Technical Symposium on Computer Science Education V. 2}{March 20--23, 2024}{Portland, OR, USA}
\acmBooktitle{Proceedings of the 55th ACM Technical Symposium on Computer Science Education V. 2 (SIGCSE 2024), March 20--23, 2024, Portland, OR, USA}
\acmDOI{10.1145/3626253.3635606}
\acmISBN{979-8-4007-0424-6/24/03}

\begin{document}


\title{Integrating Personalized Parsons Problems with Multi-Level Textual Explanations to Scaffold Code Writing}

\author{Xinying Hou}
\orcid{0000-0002-1182-5839}
\affiliation{%
  \institution{University of Michigan}
  \city{Ann Arbor}
  \state{Michigan}
  \country{USA}
}
\email{xyhou@umich.edu}

\author{Barbara J. Ericson}
\orcid{0000-0001-6881-8341}
\affiliation{%
  \institution{University of Michigan}
  \city{Ann Arbor}
  \state{Michigan}
  \country{USA}
}
\email{barbarer@umich.edu}

\author{Xu Wang}
\orcid{}
\affiliation{%
  \institution{University of Michigan}
  \city{Ann Arbor}
  \state{Michigan}
  \country{USA}
}
\email{xwanghci@umich.edu}

\begin{abstract}
Novice programmers need to write basic code as part of the learning process, but they often face difficulties. To assist struggling students, we recently implemented personalized Parsons problems, which are code puzzles where students arrange blocks of code to solve them, as pop-up scaffolding. Students found them to be more engaging and preferred them for learning, instead of simply receiving the correct answer, such as the response they might get from generative AI tools like ChatGPT. However, a drawback of using Parsons problems as scaffolding is that students may be able to put the code blocks in the correct order without fully understanding the rationale of the correct solution. As a result, the learning benefits of scaffolding are compromised. Can we improve the understanding of personalized Parsons scaffolding by providing textual code explanations? In this poster, we propose a design that incorporates multiple levels of textual explanations for the Parsons problems. This design will be used for future technical evaluations and classroom experiments. These experiments will explore the effectiveness of adding textual explanations to Parsons problems to improve instructional benefits.
\end{abstract}


\vspace{-4mm}

\maketitle
\vspace{-4mm}
\section{Introduction}
\begin{figure}[ht]
    \centering
    \includegraphics[width=0.98\linewidth]{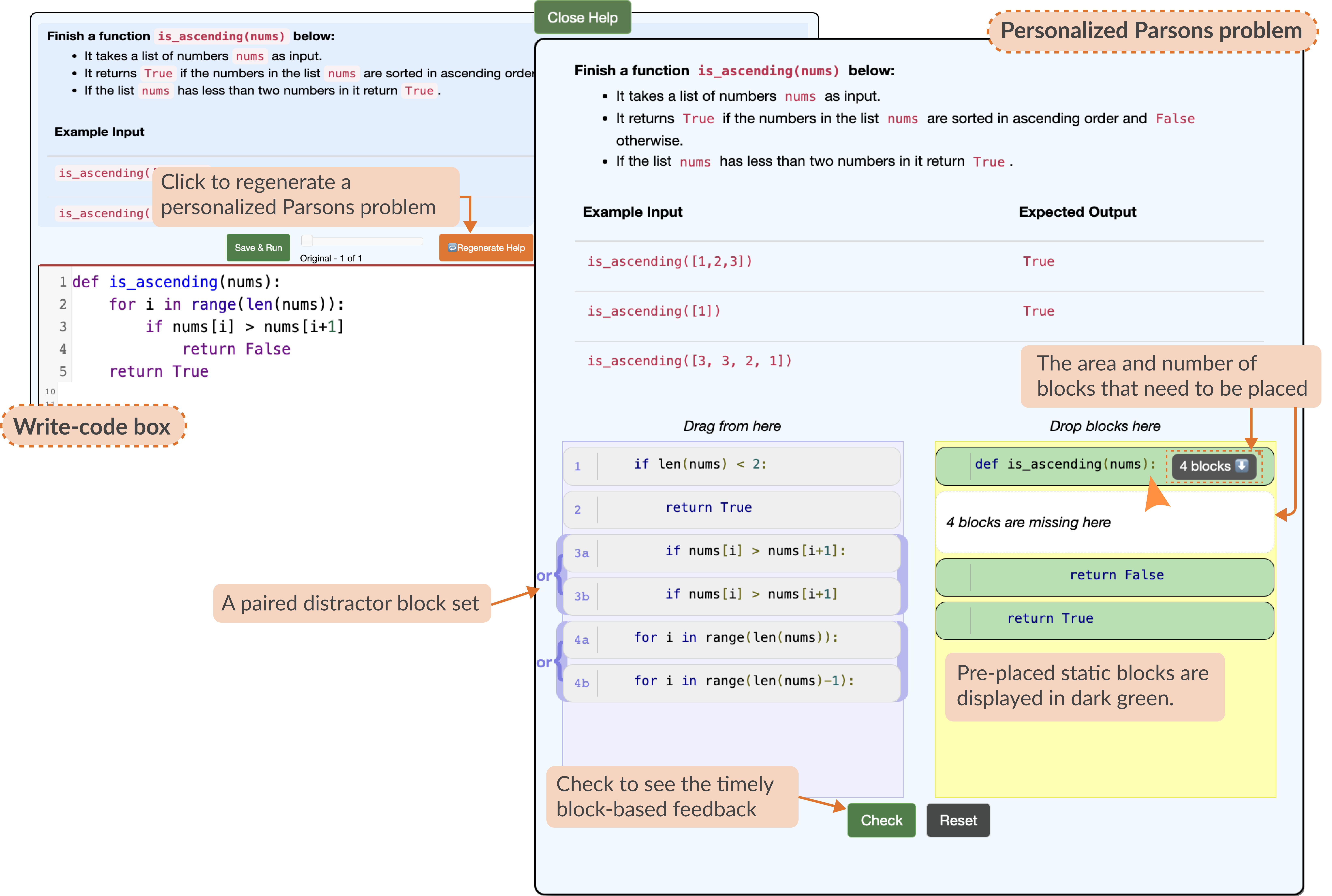}
    \caption{A write-code box (left) with a pop-up personalized Parsons problem as scaffolding (right).}

    \label{interface}
    \vspace{-4mm}
\end{figure}

One main goal of introductory programming courses is to develop basic coding skills. Although AI generation tools are now widely used in professional code development, it is still crucial for beginners to participate in code-writing activities to fully acquire fundamental programming concepts. In recent work, we explored providing Parsons problems as scaffolding for students who are struggling while writing code independently in Python \cite{hou2022using, hou2023understanding, hou2023parsons}. By leveraging the power of LLMs, we designed and implemented a system called \textit{\sys{}} to provide multi-stage personalization through a Parsons problem to assist students with code writing (Figure \ref{interface}). An initial study with 18 novice programming students showed that \sys{} is engaging and benefits learning. However, some students reported challenges as they could not fully understand the meaning of some code blocks. Textual explanations could help users understand the program's components, objectives, and structure \cite{izu2019fostering}. They can assist beginners in learning more from the Parsons problem, providing an additional opportunity to read code blocks, understand the problem, and comprehend the solution. Hence, we propose a design to incorporate multiple levels of natural language explanations into \sys{}. While explanations have been given with various programming learning materials or activities, there is limited research on integrating them with Parsons problems. Given the increasing interest and demonstrated effects of using Parsons problems to scaffold novice programmers' learning, more research is required to explore the integration of explanations into Parsons problems.
\vspace{-3mm}

\begin{figure}[ht]
    \centering
    \includegraphics[width=0.8\linewidth]{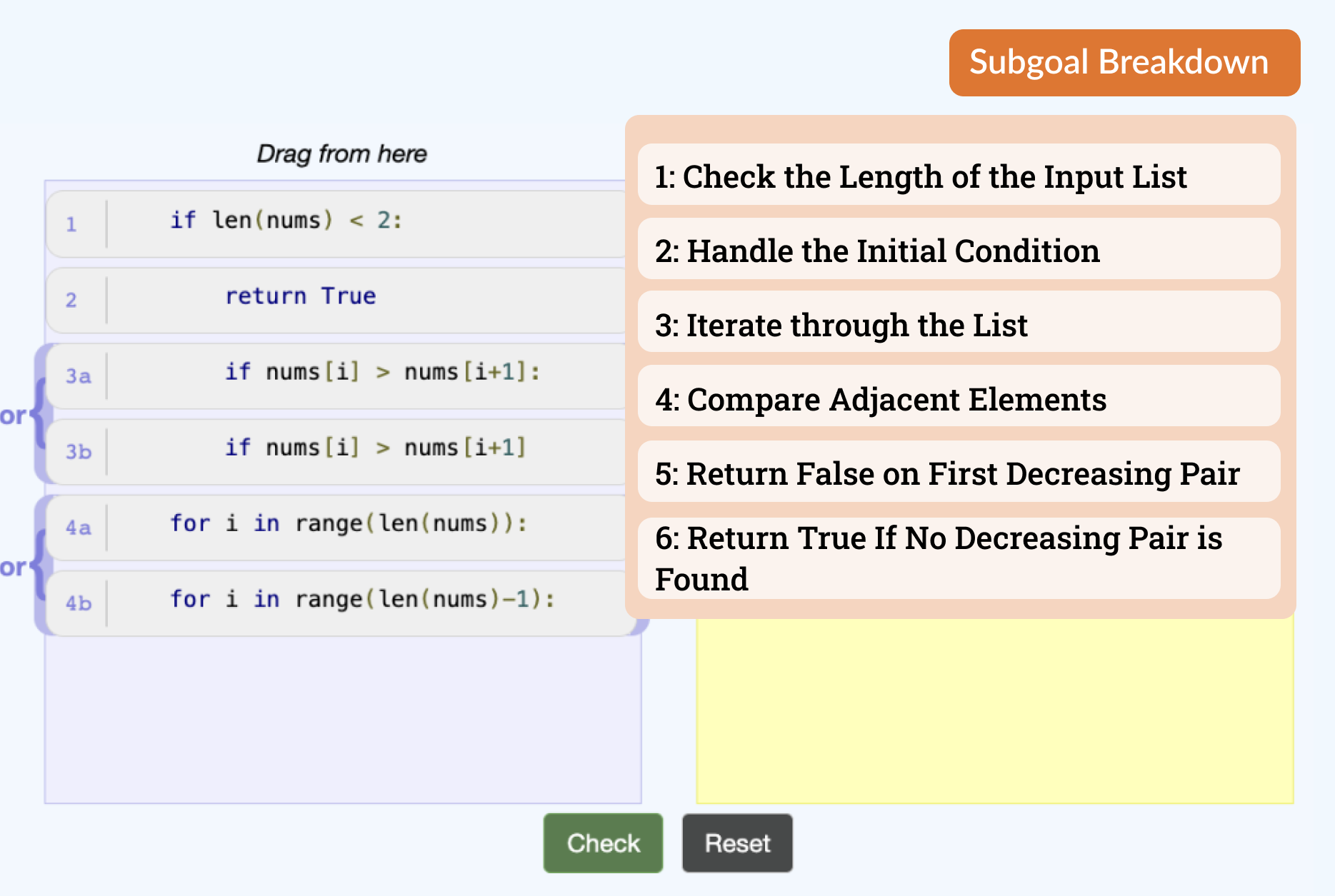}
    \caption{A subgoal list about the Parsons problem solution.}
    \label{subgoal}
    \vspace{-4mm}
\end{figure}

\begin{figure}[ht]
    \centering
    \includegraphics[width=\linewidth]{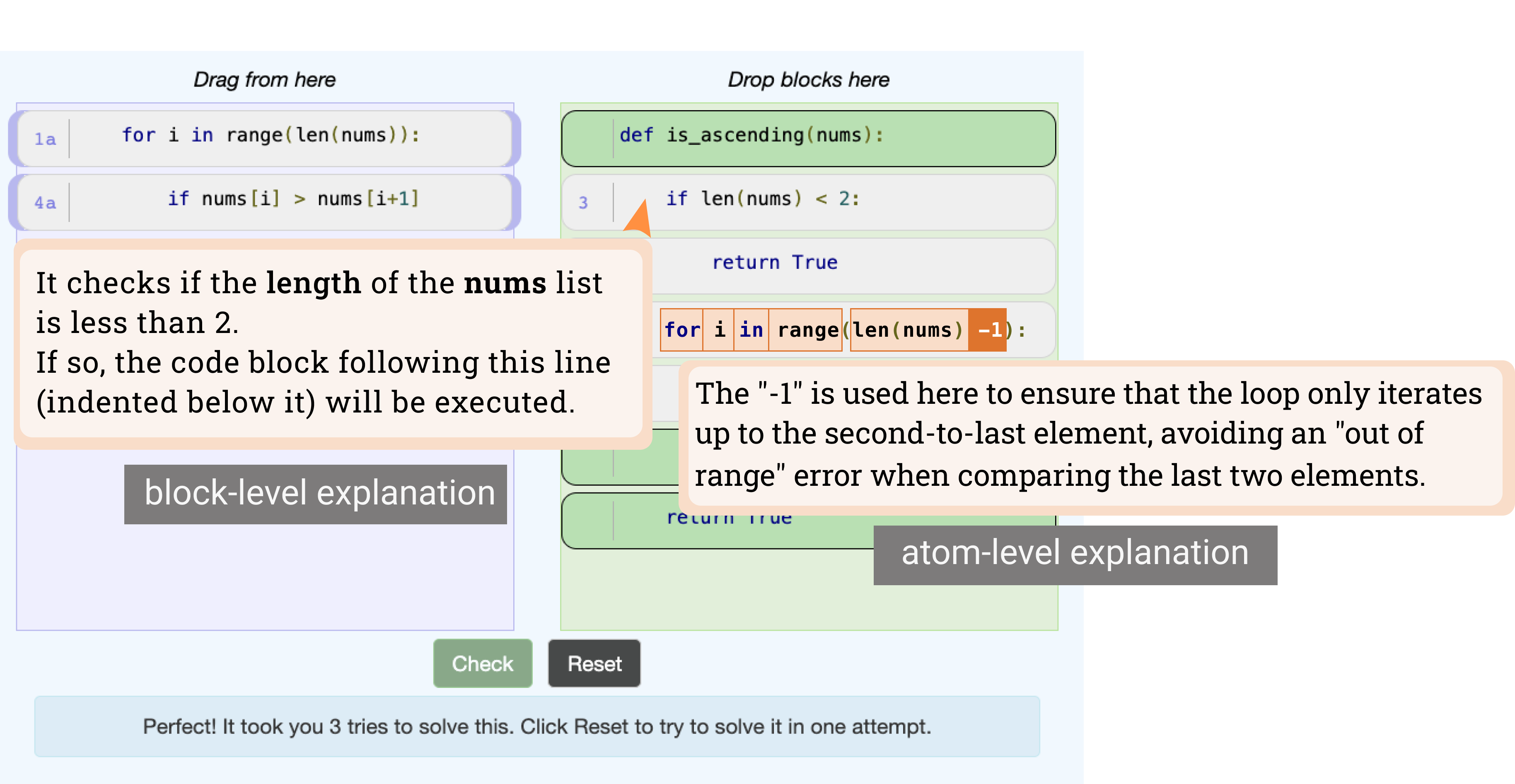}
    \caption{Block-level and atom-level explanations for the finished Parsons blocks.}
    \label{line_by_line}
    \vspace{-7mm}
\end{figure}
\vspace{-3mm}
\section{System Design}
\vspace{-0.5mm}
\sys{} is a large language model (LLM)-powered system that provides real-time, on-demand, and multi-staged personalized Parsons problems to support students while writing code. It differs from existing hint systems for programming in that \sys{} provides an active learning opportunity through the help. Specifically, students need to engage in problem-solving with the support provided rather than merely being consumers of a displayed help message \cite{chi2014icap}. An overview of \sys{}'s main interface is shown in Figure \ref{interface}. After clicking the “Help" button, \sys{} provides a personalized Parsons problem that is dynamically generated based on the student's latest code state. The personalized Parsons problem is partially complete, with the correct student-written code lines already in place and unmovable. The distractors, which are unnecessary blocks in the correct solution, are generated from the student’s own incorrect code lines. Furthermore, \sys{} adjusts the difficulty of the Parsons problem based on students' requests. If a student fails three times with a fully movable Parsons problem, \sys{} allows them to combine two code blocks into one. Once the student successfully solves the Parsons problem, they can copy the solution to the write-code box.

\textbf{We enhance \sys{} by incorporating multiple levels of LLM-generated textual explanations.} Four types of explanations will be generated and inserted in real-time, enabling \sys{} to provide personalized explanations accompanying the Parsons problem. We are still exploring methods to assess the quality of these explanations automatically.

\textbf{Subgoal guidance explanation}: Students will have access to a subgoal summary of the solution provided in the unsolved Parsons problem. This is presented in numbered bullet points, breaking down the current programming task into 4-6 more manageable task subgoals (Figure \ref{subgoal}). The subgoal guidance aims to help students gain a high-level abstract understanding and guide those who struggle to start solving a Parsons problem.

\textbf{Block-level code explanation}: After solving the Parsons problem, students can hover over each block to get an explanation of each block. These explanations clarify the behavior and purpose of the Parsons solution at the block level (Figure \ref{line_by_line}). For paired distractors, the explanation also provides reasoning for why the correct block is right and why the corresponding distractor block is incorrect.

\textbf{Atom-level code explanation}: For each block in the Parsons solution, students can also click on an atom (individual elements of the programming language, such as keywords and statements) within a code block to receive an explanation of the atom's text surface, execution (if any), and purpose (Figure \ref{line_by_line}).

\begin{figure}[ht]
    \centering
    \includegraphics[width=0.8\linewidth]{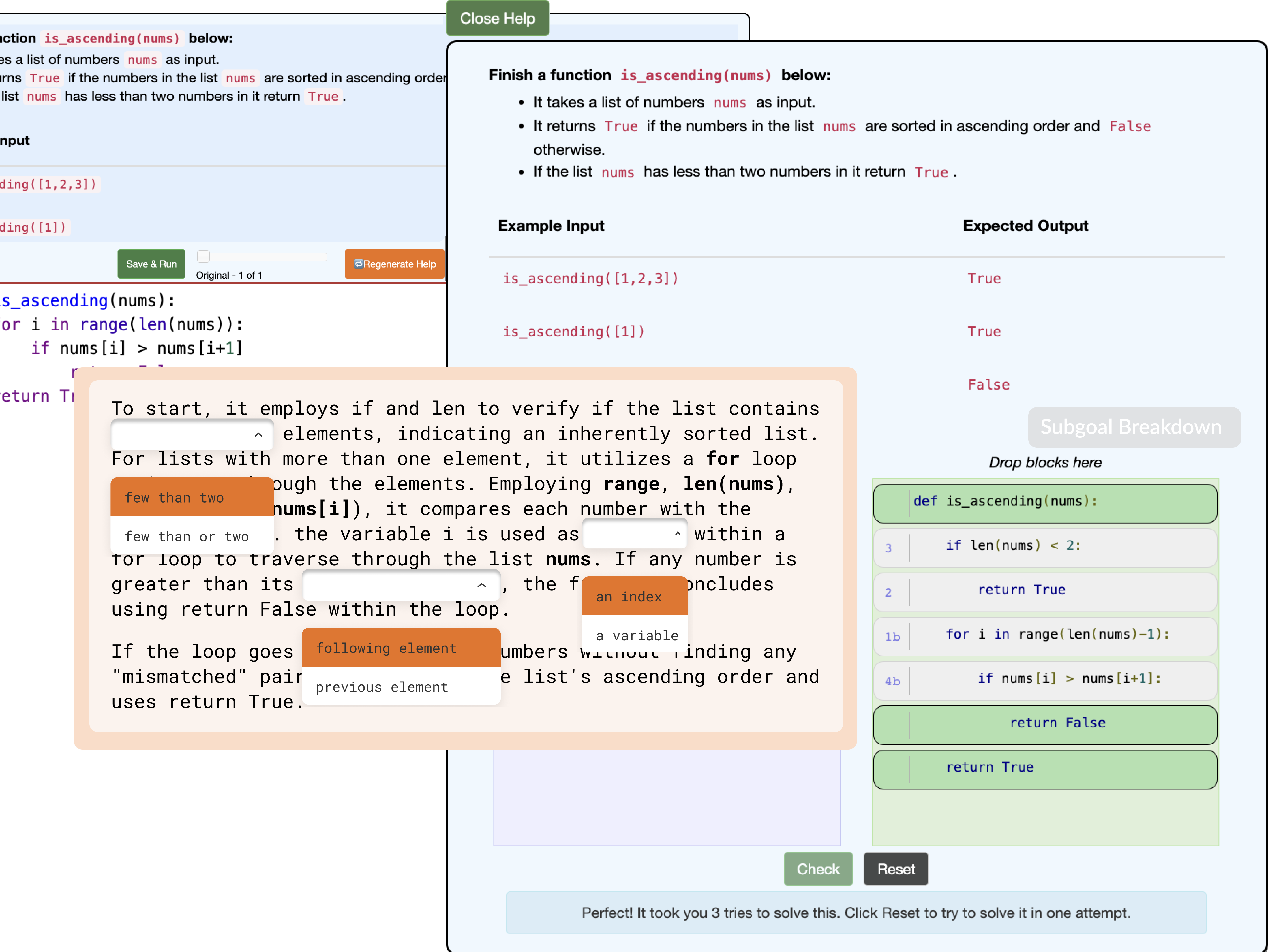}
    \caption{Students will receive a menu-based self-explanation question after solving the Parsons problem.}
    \label{self-explanation}
    \vspace{-4mm}
\end{figure}

\textbf{A self-explanation question to reflect on the Parsons solution}:
After getting the write-code problem correct, this Parsons solution will appear again without any textual explanations. This time, students need to explain the reasoning behind this solution in a menu-based self-explanation prompt. The prompt is designed to minimize guessing and enhance students' understanding. In this case, an LLM will generate the main structure of this explanation, leaving keywords to be filled in (Figure \ref{self-explanation}) by the student.
\vspace{-3mm}
\begin{acks}
The funding for this research came from the National Science Foundation award 2143028. Any opinions, findings, conclusions, or recommendations expressed in this material are those of the authors and do not necessarily reflect the views of the National Science Foundation.
\end{acks}
\vspace{-3mm}
\bibliographystyle{ACM-Reference-Format}
\bibliography{bibliography}

\end{document}